\documentclass[prl,final,twocolumn,showpacs]{revtex4}
\usepackage{graphicx}

\begin{document}
\title{Dynamical Coulomb blockade of multiple Andreev reflections}
\author{A. Levy Yeyati$^1$, J.C. Cuevas$^{1,2}$ and A. 
Mart\'{\i}n-Rodero$^1$}
\affiliation{$^1$Departamento de F\'\i sica Te\'orica de la Materia 
Condensada CV, Universidad Aut\'onoma de Madrid, E28049 Madrid, Spain}
\affiliation{$^2$Institut f\"ur Theoretische Festk\"orperphysik,
Universit\"at Karlsruhe, D-76128 Karlsruhe, Germany}

\begin{abstract}
We analyze the dynamical Coulomb blockade of multiple Andreev reflections
(MAR) in a superconducting quantum point contact coupled to a 
macroscopic impedance. We find that at very low transmission
the blockade scales as $n^2$ with $n = \mbox{Int}(2\Delta/eV)$,
where $V$ is the bias voltage and $\Delta$ is the superconducting gap,
as it would correspond to the occurrence of {\it shots} of charge $ne$.
For higher transmission the blockade is reduced both due to Pauli
principle and to elastic renormalization of the MAR probability, and
for certain voltage regions it may even become an {\it antiblockade}, i.e. 
the current is enhanced due to the coupling with the 
electromagnetic environment.
\end{abstract}


\maketitle

{\it Introduction:}
Our understanding of coherent electron transport   
in superconducting nanostructures has experienced a remarkable progress in 
the last decade \cite{superlattices}. 
A unified picture of the dc and ac Josephson effects has emerged in which 
the concept of multiple Andreev reflection \cite{KBT,tocho}  
plays a central role. 
In this respect, superconducting atomic-size contacts have provided an 
ideal system in which theoretical predictions can be tested with 
a high degree of accuracy \cite{scheer97}.
A remarkable feature of the MAR mechanism in superconducting quantum point 
contacts (SQPC) is the occurrence of coherent transfer of multiple charge 
quanta $ne$ in the subgap region with $n$ increasing as $1/V$ 
\cite{noise}. 
This is particularly striking for the case of
atomic contacts where the current can then be 
carried by the coherent transfer of several electrons through a
cross section of only few atoms.

On the other hand, the importance of charging effects in nanoscale junctions
has been stressed since the early 90's \cite{SET}. 
In these junctions with a very small 
capacitance quantum fluctuations   
result in a suppression of the current at low temperatures, 
a phenomenon known as dynamical Coulomb blockade (DCB) \cite{devoret90}. 
This effect is strongly dependent on the effective 
impedance of the circuit in which the nanoscale junction is 
embedded.
One can expect these effects to scale with the
charge of the carriers, 
which would result in a much stronger blockade when several electrons 
are transferred. 
A natural question that arises is to which extent this simple 
picture applies for the blockade of MAR in a SQPC.
Moreover, one would like to  
determine the actual signatures of DCB that could be
observed experimentally.

The aim of this Letter is to address these questions 
in the case of a single channel SQPC. For this purpose we have extended
the microscopic theory of transport in the MAR regime to take into account
charging effects and, more generally, the effect of quantum fluctuations in
the phase introduced by the electromagnetic environment.
We show that while in the tunnel limit the onset of MAR processes
is strongly blocked (the relative blockade scales as $n^2$), 
as the transmission
increases the blockade is progressively suppressed. Moreover, at high
transmission one can even observe the opposite effect, i.e. an enhancement
or {\it antiblockade} of the current in certain voltage regions
due to the coupling with the environment. 

Dynamical Coulomb blockade in a normal QPC of arbitrary transmission 
was analyzed in Refs \cite{us01,zaikin,kinderman}. 
In Ref. \cite{us01} it was shown that one can establish 
a direct link between DCB and shot noise in this type of systems. The
relation can be written in a compact way as \cite{devoret}
\begin{equation}
<\delta \hat{I}> = -\frac{1}{2e^2} \int 
d\omega J(\omega) \int d\omega' 
\mbox{sign}(\omega-\omega') \frac{\partial S_I}{\partial V}(\omega',V),
\label{supposed-relation}
\end{equation}
where $<\delta \hat{I}>$ denotes the correction in the current due to
DCB, $J(\omega)$ is the Fourier transform of the phase correlation 
function $J(t) = <\hat{\phi}(t)\hat{\phi}(0)> - <\phi^2>$, related to
the macroscopic impedance, $Z(\omega)$, characterizing the environment
by $J(t) = 2e^2/h \int d\omega \mbox{Re}Z(\omega)(e^{i\omega t}-1)/\omega$
and $S_I(\omega,V)$ 
denotes the current noise spectrum of the QPC.
In subsequent papers Kinderman et al. 
\cite{kinderman} extended the analysis of
interaction effects to the full current distribution. 
In particular, in the 
second paper of Ref.\cite{kinderman} it was shown that at zero temperature
and for a conductor with a linear current-voltage (IV) characteristic
 a relation of this type holds for the full distribution. However,
as we discuss below, this relation does not hold exactly in the case of
superconducting electrodes. The problem thus requires the use
of more general methods. 

{\it Theoretical formalism:} 
As discussed in Ref. \cite{us01}, a QPC coupled to an environment can be
described by a Hamiltonian of the form $\hat{H} =
\hat{H}_L + \hat{H}_R + \hat{H}_T + \hat{H}_{env}$ where $\hat{H}_{L,R}$
correspond to the uncoupled electrodes, which we 
describe as BCS superconductors, $\hat{H}_{env}$ to the
environment taken as a collection of bosonic modes \cite{devoret90} and
$\hat{H}_{T}=\sum_{\sigma } v \left( 
\hat{c}_{L\sigma }^{\dagger }\hat{c}_{R\sigma
} \hat{\Lambda}_{e}+\,\,h.c. \right)$
is a term coupling the leads. The translation operator 
$\hat{\Lambda}_{e}=e^{i\hat{\phi}}$ , with
$\hat{\phi}$ satisfying  $[\hat{Q},\hat{\phi}]=ie$, takes
into account the change in the charge $\hat{Q}$ of the
environment associated with the transfer process. 
Within this model the normal transmission is given by 
$\tau = 4(vW)^2/(W^2+v^2)^2$, where $1/W\pi$ is the density of states
on the normal leads at the Fermi energy \cite{tocho}; while the
current operator is given by 
$\hat{I} = \frac{ie}{\hbar} \sum_{\sigma}
v \left(\hat{c}^{\dagger}_{L\sigma}\hat{c}_{R\sigma} \hat{\Lambda}_{e} -
\mbox{h.c.}\right)$. 

The Keldysh formalism provides a general starting point to
analyze charging effects not only on the mean current but on 
all the moments of the distribution. 
The main quantity to be determined is the cumulant generating
function $S(\chi) = -\ln F(\chi)$ with \cite{Levitov}
\begin{equation}
F(\chi) =
< \hat{T}_c \exp{\left\{-\frac{i}{\hbar} \int_c \hat{H}_{T,\chi}(t) dt
\right\}} \hat{S}_c(0,t_0) > ,
\label{generating-function}
\end{equation} 
where $\hat{T}_c$ is the chronological ordering operator on the Keldysh
contour and $\hat{S}_c$ is the evolution operator of the system along
the closed time loop.
We take this contour to go from $0$ to $t_0$ and back,  
$t_0$ being
the observation time (assumed to be much larger than any other characteristic 
time). In Eq.(\ref{generating-function}) the system is coupled to
a ``counting field" $\chi(t)$, which enters as an 
additional phase factor $e^{i\chi(t)/2}$ in $\hat{H}_T$. This field
changes sign on the two branches of the Keldysh contour, i.e.
$\chi(t_\pm)= \pm \chi$.
The mean current and all the cumulants $C_n$ of the distribution are
obtained by repeated differentiation with respect to the $\chi$,
i.e $C_n = 
-(-2ie)^n \frac{\partial^n S}{\partial \chi^n}\rfloor_{\chi=0}$ 
\cite{belzig}. 

As in Refs.\cite{us01,kinderman} we assume that the external 
impedance is small ($z = Z/(h/e^2) \ll 1$) and expand the 
generating function with respect to 
$\delta \hat{H}_T = \hat{H}_T - \hat{H}^{(0)}_T$, where $\hat{H}^{(0)}_T$ 
denotes the coupling term 
in the ideal voltage biased case. To first order in $Z$
we obtain
\begin{equation}
\delta S(\chi) = -\frac{1}{e^2} \int_c dt_1
\int_c dt_2 J(t_1,t_2) K(t_1,t_2,\chi)
\label{corrected-S}
\end{equation}
where 
$K(t,t',\chi) = <\hat{T}_c \hat{I}_{\chi}(t) 
\hat{I}_{\chi}(t') e^{-\frac{i}{\hbar}\int_c \hat{H}^{(0)}_{T,\chi}(t)dt}>$.
The correction to the mean current obtained by deriving
$\delta S$ with respect to $\chi$ does not exactly
coincide with Eq.(\ref{supposed-relation}).
In the normal state, where 
the energy dependence of the transmission can in general be neglected,
both expressions are equivalent.
However, in the superconducting case
Eq.(\ref{supposed-relation}) is only valid when $S_I(\omega,V)$ is 
sufficiently smooth in the 
scale $\omega_0$ of the typical energies characterizing  
the environment, which in general requires
$\omega_0 \ll \Delta$.

Applying Wick's theorem in Eq.(\ref{corrected-S}) we obtain an expression 
for $\delta S$ in
terms of the Keldysh-Nambu Green functions 
$G^{\alpha,\beta}_{i,j}(t,t',\chi) 
= -i <\hat{T}_c \hat{\psi}_i(t_{\alpha}) \hat{\psi}^{\dagger}_j(t'_{\beta})
e^{-\frac{i}{\hbar}\int_c \hat{H}_{T,\chi}(t)dt}>$, where
$\hat{\psi}^{\dagger}_i = \left(\hat{c}^{\dagger}_{i\uparrow}, 
\hat{c}_{i\downarrow} \right)$ and $J^{\alpha,\beta}(t,t')=<\hat{T}_c
\hat{\phi}(t_{\alpha})\hat{\phi}(t'_{\beta})> -<\hat{\phi}^2>$.
The indexes $i,j\equiv \pm1$ denote the left and right electrodes
(let $L\equiv1$ and $R\equiv-1$). 
Then, Eq.(\ref{corrected-S}) can be written as
\begin{eqnarray}
\delta S(\chi) = \sum_{\alpha,\beta = \pm} \int_0^{t_0} dt_1 \int_0^{t_0} dt_2
J^{\alpha,\beta}(t_1,t_2) \, \times \nonumber \\
\sum_{i,j\equiv \pm 1} (i \cdot j) 
\mbox{Tr} \left[\hat{G}^{\alpha,\beta}_{i,j}(t_1,t_2,0) 
\hat{\sigma}_z \hat{T}^{\beta,\alpha}_{j,i}(t_2,t_1,\chi) 
\hat{\sigma}_z \right]
\end{eqnarray}
where $\hat{T}^{\alpha,\beta}_{i,j}(t_1,t_2,\chi) =
\hat{v}^{\alpha}_{i,-i}(t_1) 
\hat{G}^{\alpha,\beta}_{-i,-j}(t_1,t_2,\chi) \hat{v}^{\beta}_{-j,j}(t_2)$,
$\hat{v}^{\alpha}_{i,-i}(t)$ being a matrix in Nambu space associated
with the hopping from $i$ to $-i$:
\begin{equation}
\hat{v}^{\alpha}_{LR}(t) = \alpha v \left(\begin{array}{cc} 
e^{i\alpha \chi/2}e^{ieVt/\hbar} & 0 \\
0 & -e^{-i\alpha\chi/2} e^{-ieVt/\hbar} \end{array} \right), 
\end{equation}
with $\hat{v}^{\alpha}_{LR}(t) = 
\left(\hat{v}^{\alpha}_{RL}(t)\right)^{\dagger}$.
By means of a double Fourier transform the Green functions can be
expressed as $\hat{G}(\omega,\omega') = \sum_n \hat{G}_{0n}(\omega)
\delta(\omega-\omega'+neV)$ \cite{tocho} and thus
\begin{eqnarray}
\delta S(\chi) = t_0 \sum_{\alpha,\beta} 
\int d\omega \int d\omega' J^{\alpha,\beta}(\omega-\omega') \, \times
\nonumber\\
\sum_{i,j=\pm 1} (i \cdot j) \sum_ n \mbox{Tr}\left[
\hat{G}^{\alpha,\beta}_{0n,ij}(\omega,0)
\hat{\sigma}_z \hat{T}^{\beta,\alpha}_{n0,ij}(\omega',\chi) 
\hat{\sigma}_z \right]
\label{deltaS-GFs}.
\end{eqnarray}

To determine the Green functions entering in the evaluation
of $\delta S(\chi)$ one has to solve the corresponding Dyson equation
$\check{\bf G} = \check{\bf g} +
\check{\bf g} \circ \check{\bf v} \circ
\check{\bf G}$, where the $\circ$ product is a shorthand for summation
over intermediate indexes and $\check{\bf g}$
are the unperturbed Keldysh Green functions
of the uncoupled electrodes
in equilibrium
\begin{equation}
\check{g}_{ij} = \delta_{ij} 
\left(\begin{array}{cc} n\hat{g}^a + (1-n) \hat{g}^r &
n \left(\hat{g}^a -\hat{g}^r\right) \\
(n-1) \left(\hat{g}^a -\hat{g}^r\right) & -n\hat{g}^r - (1-n) \hat{g}^a
\end{array} \right),
\end{equation}
where $\hat{g}^a = \hat{g}^{r,*} = g \hat{I} + f \hat{\sigma}_x$,
with $g = -\omega/W\sqrt{\Delta^2 - \omega^2} = (-\omega/\Delta) f$,
are the advanced and retarded BCS Green functions, and $n(\omega)$ 
is the Fermi factor.

\begin{figure}
\includegraphics[width=0.9\columnwidth]{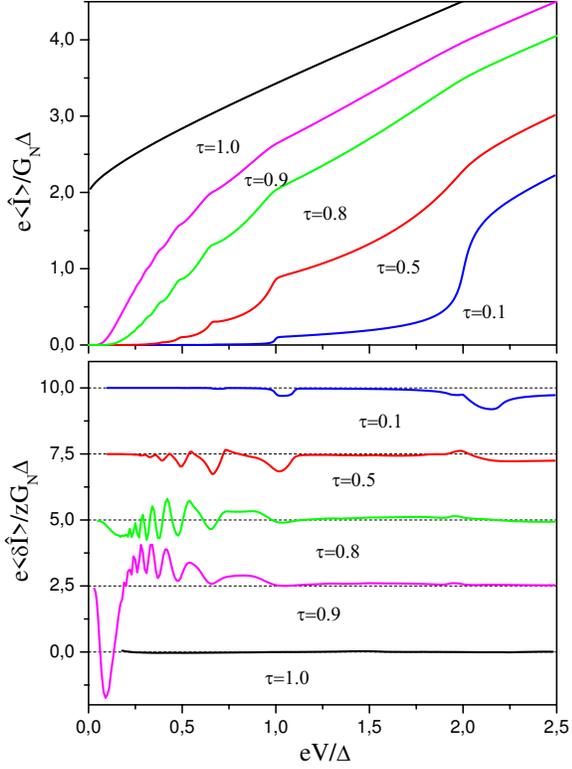}
\caption{(color online)
Mean current (upper panel) and current blockade (lower panel)
for a single mode environment with $\omega_0 = 0.2 \Delta$ at zero temperature 
for different transmission values. The
current blockade curves have been displaced vertically for clarity.
Notice normalization with $G_N= (2e^2/h)\tau$ and
$z$, which is assumed to be a small parameter.} 
\label{global}
\end{figure}

{\it Results:} 
In order to illustrate its effect on the IV characteristic
 we shall consider the simplest case of an environment characterized
by a single mode of frequency $\omega_0$.
A more general situation can be straightforwardly 
analyzed as a superposition of 
modes weighted by $\mbox{Re}Z(\omega)$. 
Fig. 1 shows the overall behavior of the current blockade at
zero temperature 
obtained by numerical evaluation of Eq.(\ref{deltaS-GFs})
for $\omega_0 = 0.2 \Delta$ and different 
values of the transmission. We also show for comparison the 
mean current $<\hat{I}>$ in the absence of environmental effects.
The most prominent features are: i) in the tunnel limit the blockade
appears around the threshold MAR voltages $2\Delta/n$,
ii) as the transmission increases the region where blockade is 
effective moves towards $V\rightarrow0$, vanishing eventually
for perfect transmission and
iii) for high transmission and
low bias the correction to the current $<\delta\hat{I}>$ can
exhibit a sign change. 
These features are discussed in more detail below.

Let us first discuss the $\tau \rightarrow 0$ regime.
Here, at a fixed bias voltage $V$, $\delta S$ is mainly determined
by processes of order $n \sim 2\Delta/eV$ in the transmission.
These processes are illustrated by the different diagrams 
in Fig. 2, which correspond to $n=3$.
Diagrams of type a) in Fig. 2 correspond to a real excitation    
of the environmental mode which gives a maximum contribution to 
$\delta S$ in the voltage range $2\Delta/n \le eV \le (2\Delta+\omega_0)/n$, 
i.e. between the opening of a MAR of order $n$ 
until the onset of inelastic  processes. More precisely, their
contribution to $\delta S$ at zero temperature
around $eV \sim 2 \Delta/n$ is given by
\begin{eqnarray}
\delta S^{(a)}(\chi) \simeq -\frac{ze\pi^2}{h}
\left(\frac{v}{4}\right)^{2n} e^{-in\chi}  \, \times \nonumber\\
\left\{ \int_{\Delta-neV}^{-\Delta} d\omega \rho(\omega) \rho(\omega+neV)
K_n(\omega,\omega)  \right. \, - \nonumber \\
\left. \int_{\Delta-neV}^{-\Delta-\omega_0} d\omega 
\rho(\omega+\omega_0) \rho(\omega+neV) 
K_n(\omega+\omega_0,\omega) 
\right\}
\label{deltaS-tunnel}
\end{eqnarray}
with 
\begin{eqnarray}
K_n(\omega,\omega') =
\left|\sum_{j=1}^{n} \prod_{k=1}^{n-j} f(\omega+keV)
\prod_{l=n-j+1}^{n-1} f(\omega'+leV) \right|^2, \nonumber
\end{eqnarray}
where $\rho(\omega) = \mbox{Im} g(\omega)/\pi$ is the BCS density of states.
On the other hand, diagrams of type b) in Fig. 2 correspond to an
elastic renormalization of the tunneling rates. In the tunnel
limit they give a much smaller contribution which
can be neglected for $\omega_0 \ll \Delta$.  
From Eq. (\ref{deltaS-tunnel}) and taking into account that
in this limit $<\hat{I}> \simeq \frac{e\pi^2}{h}
\left(\frac{v}{4}\right)^2n 
\int_{\Delta-neV}^{-\Delta} d\omega \rho(\omega) \rho(\omega+neV)
\Gamma_n(\omega)$, where $\Gamma_n(\omega) =
\prod_{k=1}^{n-1}\left|f(\omega+keV)\right|^2$ \cite{tocho}, 
we find that the relative blockade 
$-<\delta \hat{I}>/z<\hat{I}>$ reaches its maximum value $n^2$
for $2\Delta/n \le V \le (2\Delta+\omega_0)/n$.  
This result is consistent with the fact that in the
tunnel limit the transmitted charge is well defined 
and increases in a staircase way as the voltage is reduced 
\cite{noise}. In this limit the environment produces a blockade 
of MAR processes as if they would correspond to single "shots" of
charge $ne$ until their energy is 
sufficiently large to excite an environmental mode. 

\begin{figure}
\includegraphics[width=0.9\columnwidth]{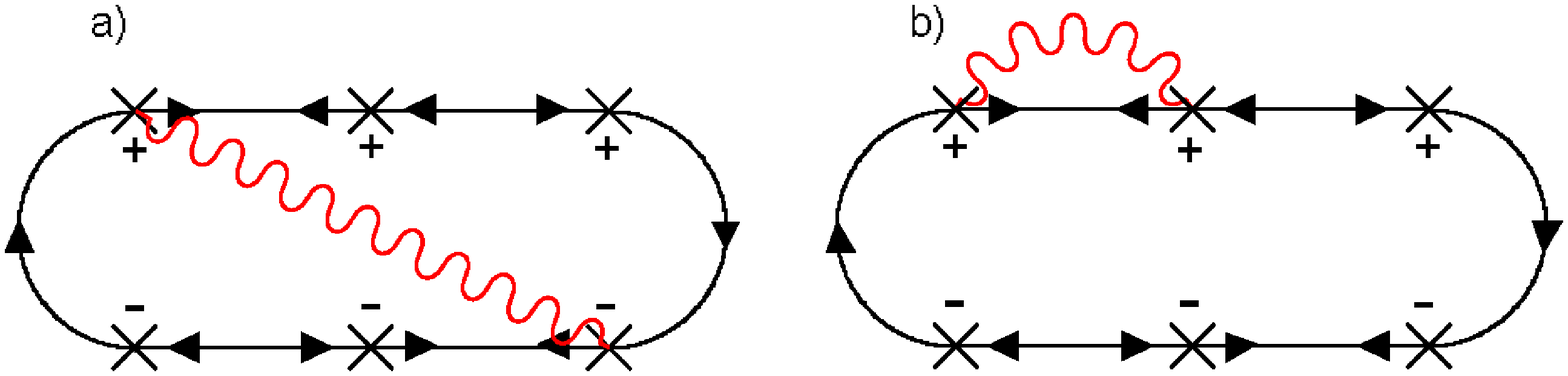}
\caption{
(color online)
Typical diagrams contributing to the blockade in the tunnel limit 
$eV \sim 2\Delta/3$. 
Crosses indicate the hopping events on the two
branches (+ or -) of the Keldysh contour. 
Full lines with an arrow indicate electron or
hole propagators, while double arrowed lines correspond to the anomalous 
ones and wavy lines denote phase correlators. 
Diagram a)
is an inelastic correction due to the coupling with the environment 
while diagram b) corresponds to an elastic renormalization of the
MAR probability.}
\label{diagrams}
\end{figure}

There are, however, {\it two} basic mechanisms which 
reduce the blockade at finite  
transmissions and finite $\omega_0$.
First, elastic renormalization of the tunneling rates 
(diagrams b) in Fig. 2) give a contribution to $\delta S$
which in the tunnel limit and for small frequencies
can be written as
\begin{eqnarray}
\delta S^{(b)}(\chi) \simeq  \frac{ze\pi^2}{h}
\left(\frac{v}{4}\right)^{2n}
\omega_0^2 (n-1) e^{-in\chi} \nonumber\\
\left\{\int_{\Delta-neV}^{-\Delta} d\omega \rho(\omega) \rho(\omega+neV)
\frac{\partial^2 \Gamma_n(\omega)}{\partial \omega^2} 
\right\} .  
\end{eqnarray}
At the opening of the Andreev channels 
$\Gamma_n$ exhibits a minimum as a function of energy and thus
the contribution to the current arising from 
$\delta S^{(b)}$ is positive. 
On the other hand, as the transmission increases there is a 
suppression of blockade at large bias due to Pauli principle which is closely
related to the reduction observed in shot noise \cite{noise}. 

\begin{figure}
\includegraphics[width=0.9\columnwidth]{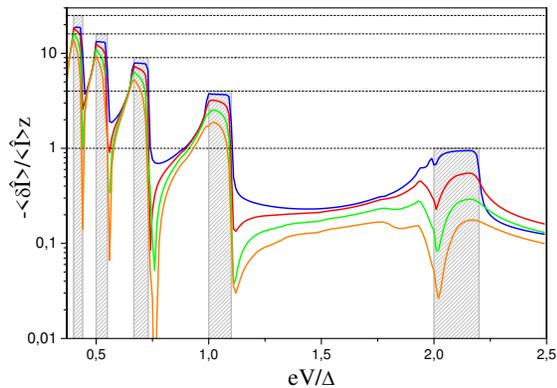}
\caption{
(color online)
Relative blockade for $\tau =$ 0.001, 0.1, 0.2, 
and 0.4 (from top to bottom) and $\omega_0 = 0.2\Delta$. 
The shadowed regions indicate the maximum relative blockade 
(scaling as $n^2$) which is predicted for $\tau \rightarrow 0$.} 
\label{ratio}
\end{figure}

The evolution of the relative
blockade for $\tau$ ranging from 0.01 to 0.4 is 
illustrated in Fig. \ref{ratio}.
The maximum relative blockade expected in the tunnel limit 
is indicated by the shadowed regions. As can be observed 
deviations from the $n^2$ scaling are already significant for
transmissions of the order of 0.1.  

\begin{figure}
\includegraphics[width=0.9\columnwidth]{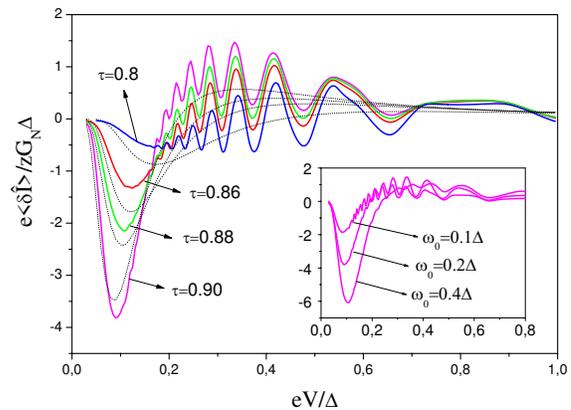}
\caption{
(color online)
Current blockade at small bias voltage and large transmission. 
Notice the sign change for intermediate bias. The dotted lines correspond
to the approximate expression of Eq.(\ref{aprox-LZ}). 
The inset shows the behavior
of blockade for $\tau=0.9$ and different values of $\omega_0$.}
\label{high-trans}
\end{figure}

The behavior of the blockade is qualitatively different for higher
transmissions.
This is illustrated in Fig. \ref{high-trans}, which shows $<\delta \hat{I}>$ 
close to the ballistic limit and for low bias voltage.
It is observed that a blockade peak at low
bias is followed by a region where the system exhibits antiblockade.
The overall behavior is robust with respect 
to changes in $\omega_0$ from 0 to $\sim \Delta$ 
(see inset in Fig. \ref{high-trans}).  
One can understand this behavior from Eq.(\ref{supposed-relation})
which is valid when $\omega_0 \rightarrow 0$. In this limit
and at zero temperature we have $<\delta \hat{I}> \propto -\omega_0 
\frac{\partial S(0)}{\partial V}$. 
By virtue of this relation 
the blockade can be interpreted as a {\it reaction} of the system
in order to compensate the power which is dissipated in the circuit
by the current fluctuations introduced by the mesoscopic conductor. 
In a SQPC at high transmission the noise exhibits regions with negative slope 
($\partial S_I/\partial V < 0$) \cite{noise}. 
In these regions the compensation
requires an enhancement of the current instead of a reduction.   
This behavior can be obtained in a simple way as follows.
Close to the ballistic limit the noise at low bias arises from
Landau Zener transitions between Andreev states 
(see second Ref. in \cite{noise}), and is given
by $S(0) = \frac{2\Delta}{V} p \left(1-p\right)$, where 
$p = \exp{\left[-\pi\Delta(1-\tau)/eV\right]}$ is the Landau-Zener 
probability. As a result one obtains
\begin{equation}
<\delta \hat{I}> \sim \omega_0 \frac{2\Delta}{V^2} p\left[\left(1-p
\right) - \frac{\pi\Delta(1-\tau)}{eV} \left(1-2p\right)\right] ,
\label{aprox-LZ}
\end{equation} 
which exhibits a sign change for $eV \simeq 1.75 \Delta (1-\tau)$.

{\it Conclusions:}
We have analyzed dynamical Coulomb blockade in a SQPC in series
with a macroscopic impedance. A scaling with the square of the effective
charge is found in the limit of vanishing transmission and 
environmental
frequencies. As the transmission increases the blockade is strongly
suppressed for $eV \sim \Delta$ and for certain voltage range it may even
become an {\it antiblockade}. Our predictions could be tested experimentally
in superconducting atomic contacts with a well characterized electromagnetic
environment, like those already used 
to analyze DCB in the normal state \cite{cron01}.

\acknowledgements
The authors would like to thank fruitful discussions with M.H. Devoret, 
D. Stevy and C. Urbina.
JCC was financially supported by the Helmholtz Gemeinschaft. 


\begin{thebibliography}{}
\bibitem{superlattices} Special issue Superlattices and Microstruct.
{\bf 25}, No. 5/6 (1999).
\bibitem{KBT} T.M. Klapwijk, G.E. Blonder, and M. Tinkham, Physica 
(Amsterdam) {\bf 109B\&110B}, 1657 (1982); G.B. Arnold, J. Low Temp. Phys. 
{\bf 68}, 1 (1987); E. N. Bratus, V.S. Shumeiko and G. Wendin, Phys. Rev. Lett.
{\bf 74}, 2110 (1995); D. Averin and A. Bardas, Phys. Rev. Lett. {\bf 75}, 1831
(1995).
\bibitem{tocho} J.C. Cuevas, A. Mart\'{\i}n-Rodero and A. Levy Yeyati,
Phys. Rev. B {\bf 54}, 7366 (1996).
\bibitem{scheer97} E. Scheer {\it et al.}, Phys. Rev. Lett. {\bf 78}, 
3535 (1997); E. Scheer {\it et al.}, Nature {\bf 394}, 154 (1998); 
M. Goffman {\it et al.}, Phys. Rev. Lett. {\bf 85}, 170 (2000); 
B. Ludoph {\it et al.} Phys. Rev. B {\bf 61}, 8561 (2000);
R. Cron {\it et al.} Phys. Rev. Lett. {\bf 86}, 4104 (2001).
\bibitem{noise} J.C. Cuevas, A. Mart\'{\i}n-Rodero and A. Levy Yeyati,
Phys. Rev. Lett. {\bf 82}, 4086 (1999); Y. Naveh and D.V. Averin, Phys. 
Rev. Lett. {\bf 82}, 4090 (1999); J.C. Cuevas and W. Belzig, Phys. Rev.
Lett. {\bf 91}, 187001 (2003); G. Johansson, P. Samuelsson, and A. Ingerman
Phys. Rev. Lett. {\bf 91}, 187002 (2003)
\bibitem{SET} {\it Single Charge
Tunneling}, edited by H. Grabert and M.N. Devoret (Plenum Press, New York,
1992).
\bibitem{devoret90}  M.H. Devoret et al., Phys. Rev. Lett. {\bf 64}, 1824
(1990); S.M. Girvin et al., Phys. Rev. Lett. {\bf 64}, 3183 (1990).
\bibitem{us01} A. Levy Yeyati {\it et al.}, Phys. Rev. Lett. {\bf 87},
046802 (2001).
\bibitem{zaikin} D.S. Golubev and A.D. Zaikin, 
Phys. Rev. Lett. {\bf 86}, 4887 (2001).
\bibitem{kinderman} M. Kindermann and Yu. V. Nazarov
Phys. Rev. Lett. {\bf 91}, 136802 (2003); M. Kindermann, Yu.V. Nazarov, 
and C.W.J. Beenakker, Phys. Rev. B {\bf 69}, 035336 (2004).
\bibitem{devoret} M.H. Devoret (private communication).
\bibitem{Levitov} L.S. Levitov in {\it Quantum Noise in Mesoscopic
Physics}, edited by Yu.V. Nazarov, Kluwer Academic Press (2002), pag. 373.
\bibitem{belzig} W. Belzig {\it ibidem}, pag. 463
\bibitem{cron01} R. Cron et al. in {\it Electron Correlations
from Meso to Nano-Physics}, edited by T. Martin, G. Montambaux and
J. Tr{\^a}n Thanh V{\^a}n, Les Ulis, France (2001), pag. 17.
\end{thebibliography}
\end{document}